\documentclass[graybox]{svmult}

\usepackage{mathptmx}       
\usepackage{helvet}         
\usepackage{courier}        
\usepackage{type1cm}        
\usepackage{makeidx}         
\usepackage{graphicx}        
\usepackage{multicol}        
\usepackage[bottom]{footmisc}

\usepackage{verbatim,color}
\usepackage{amsmath}
\usepackage{subfigure}
\usepackage{setspace}
\usepackage{url}
\usepackage{graphicx}
\usepackage{graphics}
\usepackage{enumerate}
\usepackage{framed, color}
\usepackage{framed}
\usepackage{caption}
\usepackage{verbatim}
\makeatletter
\newcommand{\verbatimfont}[1]{\def\verbatim@font{#1}}%
\makeatother

\makeindex             

\begin{document}

\title*{Physicists' approach to studying socio-economic inequalities: Can humans be modelled as atoms?}
\titlerunning{Studying socio-economic inequalities}
%
%
\author{Kiran Sharma and Anirban Chakraborti}

\institute{ Kiran Sharma \at School of Computational and Integrative Sciences, Jawaharlal Nehru University, New Delhi-110067, India, \email{kiran34_sit@jnu.ac.in}
\and Anirban Chakraborti \at School of Computational and Integrative Sciences, Jawaharlal Nehru University, New Delhi-110067, India, \email{anirban@jnu.ac.in}}

%
\maketitle

\abstract{ A brief overview of the models and data analyses of income, wealth, consumption distributions by the physicists, are presented here. It has been found empirically that the distributions of income and wealth possess fairly robust features, like the bulk of both the income and wealth distributions seem to reasonably fit both the log-normal and Gamma distributions, while the tail of the distribution fits well to a power law (as first observed by sociologist Pareto). We also present our recent studies of the unit-level expenditure on consumption across multiple countries and multiple years, where it was found that there exist invariant features of consumption distribution: the bulk is log-normally distributed, followed by a power law tail at the limit. The mechanisms leading to such inequalities and invariant features for the distributions of socio-economic variables are not well-understood. We also present some simple models from physics and demonstrate how they can be used to explain some of these findings and their consequences. } 
\section{Introduction}
\label{intro}

Physicists have been always keen on exploring domains outside of physics, like biology, geology, astronomy, sociology, economics, etc., often giving birth to very successful interdisciplinary subjects like biophysics, astrophysics, geophysics, sociophysics, econophysics and so on \cite{Chakraborti2016}. The last two interdisciplinary fields: Sociophysics \cite{Sen2013,Chakrabarti2006} and Econophysics \cite{Sinha2010,Slanina2013}, have been only recent additions to the long list. However, the physicists’ interest in the social sciences (Economics and Sociology) is quite old, and they have been trying to approach economic and social problems using their experience of modelling physical systems and analysing data. The physicists' ability to deal with complex dynamical systems have often inspired the central ideas and foundations of the modern axiomatic foundations of economics. No wonder, the first Nobel Prize winner in economics was Jan Tinbergen, a physicist by training (having completed his Ph.D. from the University of Leiden in 1929 on `Minimisation problems in Physics and Economics') \cite{Samuelson2016}. The book of Paul Samuelson, the second Nobel Laureate in economics, entitled “Foundations of Economic Analysis” (1947) \cite{Samuelson1947}, which is considered as his magnum opus — derived from his doctoral dissertation at Harvard University, makes use of the classical thermodynamic methods of Willard Gibbs \cite{wikiGibbs}, the American physicist and one of the founders of the area of statistical physics. On the other hand, the economic concept of the `invisible hand' due to Adam Smith \cite{wikiSmith}, cited as the father of modern economics and best known for his two classic works: The Theory of Moral Sentiments (1759) \cite{Smith1759}, and An Inquiry into the Nature and Causes of the Wealth of Nations (1776) \cite{Smith1776}, can be understood as an attempt to describe the influence of the market as a spontaneous order on people's actions and self-organization, which has influenced many models of physical systems. Many such cross-fertilisation of ideas and concepts have been taking place for a long time, eventually leading to intense activities in the field and the coinage of the word:

\vspace{0.3cm}

\noindent\fbox{

\parbox[c]{\textwidth}{
\textbf{Econophysics}, a new interdisciplinary research field applying methods of statistical physics to problems in economics and finance, first introduced by the theoretical physicist H. Eugene Stanley in an international conference on statistical physics at Kolkata (India) in 1995.}
}
\vspace{0.3cm}
  
In the field of statistical physics \cite{Mandl2002,Haar2006,Sethna2006}, one often encounters a system of many interacting dynamical units exhibiting a `collective behaviour', which simply depends on a few basic (dynamical) properties of the individual constituents and the embedding dimension of the system. Since it is independent of other details, it thus displays a sort of `universality'. Often, socio-economic data also exhibit enough empirical evidences in support of such `universalities', which prompt the physicists' to propose simple, minimalistic models to understand them using the methods of statistical physics.

In sociology and economics, some of the major issues of concern have beeb inequalities in different forms: income, wealth, etc. It has been argued by certain social scientists that a society or country performs better where the resources are distributed more equitably or there exists less inequalities between the haves and the have-nots \cite{Wilkinson2009}. Most economists agree that fairness or equal chances of being involved in the economic activities promote growth \cite{Rodrik2003}. It would be impossible to find any society or country where e.g., income or wealth, is equally distributed among its people. The distribution of wealth, income, and consumption has never been uniform, and economists (and very recently physicists) have tried for years to understand the reasons for such inequalities. For a very long time, scholars have been working on the statistical descriptions and mechanisms leading to such inequalities (see e.g., Refs.~\cite{Sen1992,BKChakrabarti2013,Deaton1992}). We briefly mention a few of the empirical observations in the first section of the article. Based on these observations, many questions have been formulated by different scholars. In the following sections, we address from the perspective of the physicists, the questions below:
\begin{itemize}
\item	How are income, wealth, and consumption distributed and what are the statistical forms of their distributions?
\item	Are there any robust or ``universal'' features of the statistical forms and how can they be modelled/reproduced in a mathematical/computational framework?
\end{itemize}

\section{Empirical distributions of income and expenditure}	
Following several studies spanning more than a century, a few established regularities in income and wealth distributions have been observed. The most popular regularity was proposed by the Italian sociologist and economist, Vilfredo Pareto, who made extensive studies at the end of the 19th century and found that wealth distribution in Europe follows a power law for the very rich \cite{Pareto1897}. Later this came to be known as the Pareto law. Subsequent studies revealed that the distributions of income and wealth possess other fairly robust features, like the bulk of both the income and wealth distributions seem to reasonably fit both the log-normal and the Gamma distributions, also sometimes known as Gibrat’s law \cite{Gibrat1931}. 
    \begin{framed}
    \noindent \textbf{Pareto Law:} In 1897, Pareto made extensive studies in Europe and found that wealth distribution follows a power law tail for richer sections of society. 
     For about $90-95\%$ of the population, the distribution matches a Gibbs or Gamma (black curve), while the income for the top $5-10\%$ of the population decays much more slowly, following a power- law (red line). 
           \includegraphics[width=4cm]{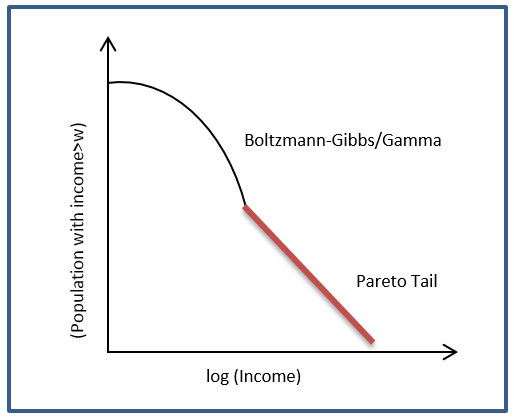}%
    \captionof{figure}{\textbf{Plot for the cumulative income distribution.} The population fraction having an income greater than a value $w$ plotted against $w$) shown on a double logarithmic scale. The exponent of the Pareto tail, known as the Pareto exponent, is given by the slope of the red line in the double-logarithmic scale, and is found to be between $1$ and $3$ (see refs.~\cite{BKChakrabarti2013,Pareto1897}).}
    \end{framed}

\begin{figure}
\begin{center}
\includegraphics[width=0.9\linewidth]{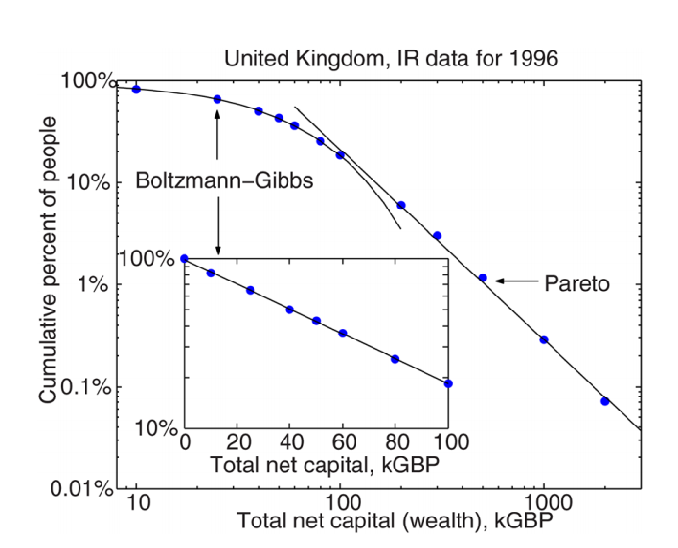}
\caption{ \textbf{Cumulative probability distribution of the net wealth.} The wealth composed of assets (including cash, stocks, property, and household goods) and liabilities (including mortgages and other debts) in the United Kingdom shown on double-logarithmic (main panel) and log-linear (inset) scales. Points represent the data from the Inland Revenue, and solid lines are fits to the Boltzmann-Gibbs (exponential) and Pareto (power) distributions (taken from Ref. \cite{Dragulescu2001}).}
\label{Fig:wealth_data}
\end{center}
\end{figure}

Physicists use the Gamma distribution for fitting the probability density or the Boltzmann-Gibbs/exponential distribution for the corresponding cumulative distribution. However, the tail of the distribution fits well to a power law (as first observed by Pareto), the exponent known as the Pareto exponent, usually ranging between $1$ and $3$ \cite{BKChakrabarti2013}. For India too, the wealthiest have been found to have their assets distributed along a power law tail \cite{Sinha2006}. 
The shape of the typical wealth distribution is thus, with Gibbs/Gamma behavior at lower and intermediate values of wealth $w < w_c$, and a Pareto (power--law) tail at the larger values, $w \geq w_c$   where $w_c $  is a crossover value that depends on the numerical fitting of the data:

\begin{eqnarray}
P(w) &\sim & w^n exp(-w/T), \hspace{1.0cm}	{\rm for} \hspace{0.5cm}	w < w_c, \nonumber \\ 
     &\sim & w^ {-\alpha-1}, \hspace{2.1cm} {\rm for} \hspace{0.5cm} w\geq w_c,
     \label{Eq:energy_dist}
\end{eqnarray}
     			     			
\noindent where $P(w)$, the equilibrium distribution of wealth, is defined as follows: $P(w)dw$ is the probability that in the steady state of the system, a randomly chosen agent will be found to have wealth between $w$ and $w+dw$. The exponent $\alpha$ is known as the Pareto exponent, as mentioned earlier, $T$ is the average wealth (analogous to the temperature in a gas) of the economic system and $n$ is a numerical constant. Detailed empirical results in support of the above statistical form for both income and wealth in different countries, economic societies and over different periods of time, can be found in several research articles, monographs and books that are mentioned in the list of references. For an illustration, Fig. \ref{Fig:wealth_data} shows the cumulative probability distribution of the net wealth, composed of
assets (including cash, stocks, property, and household goods) and liabilities
(including mortgages and other debts) in the United Kingdom for the year 1996.

\begin{figure}
\begin{center}
\includegraphics[width=0.9\linewidth]{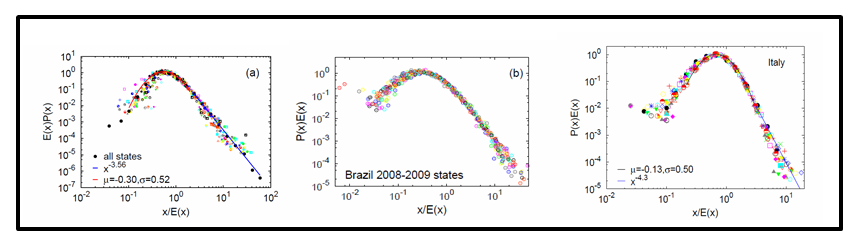}
\caption[Consumption]{\textbf{Plot for consumption expenditure data.} 
(a) For Indian states, data normalized for all states and fitted with a log--normal distribution and a power law at the right tail during the period 2011-2012.
(b) For Brazilian states, data normalized with respect to the respective mean expenditure across states during the period 2008-2009.
(c) For all states of Italy, performed during the period 1980-2012.
(Taken from Ref. \cite{Chakrabarti}).}
\label{Fig:consumption}
\end{center}

\end{figure}

Although income and wealth distribution data are mainly used to quantify economic inequality for individuals or family/households, the distribution of consumer expenditure does also reflect certain aspects of disparity in the society. In a recent study, Chakrabarti et al. \cite{Chakrabarti} analysed the unit-level expenditure on consumption across multiple countries and multiple years, and showed that certain invariant features of the consumption distribution could be extracted. Specifically, it was shown that the bulk of the distribution follows a log-normal, followed by a power law tail. As shown in Fig. \ref{Fig:consumption}, the distributions coincide with each other under normalization by mean expenditure and log scaling, even though the data was sampled across multiple dimensions including, e.g., time, social structure and locations across the globe. This observation seems to indicate that the dispersion in consumption expenditure across various social and economic groups are significantly similar (`universal'), subject to suitable scaling and normalization. 
In another article, Chatterjee et al. \cite{Chatterjee2016} studied the distributional features and inequality of consumption expenditure, specifically across India -- for different states, castes, religion and urban-rural divide. Once again they found that even though the aggregate measures of inequality are fairly diversified across the Indian states, the consumption distributions show near identical statistics after proper normalization. This feature was again seen to be robust with respect to variations in sociological and economic factors. They also showed that state-wise inequality seems to be positively correlated with growth, which is in agreement with the traditional idea of first part of the Kuznets’ curve \cite{Kuznets1955}.

Having discussed briefly the first question mentioned in the introduction, we now turn our attention to the modeling of the different robust features of the empirical distributions, by using a mathematical/computational framework that is inspired by some simple models of statistical physics of ideal gas.

\section{Kinetic exchange models} 
The simple yet powerful framework of kinetic theory of ideal gases, first proposed in 1738 by Bernoulli, led eventually to the successful development of statistical physics towards the end of the 19th century \cite{Mandl2002,Haar2006,Sethna2006}. The aim of statistical physics is to study the physical properties of macroscopic systems consisting of a large number of constituent particles. In such large systems, the number of particles is of the order of Avogadro number and it is extremely difficult to have complete microscopic description of such a system. 

The basic concept of kinetic exchange model is taken from the `\textit{Kinetic theory of gases}', which describes a gas as a large number of sub-microscopic particles (atoms and molecules), all of which are in constant, random motion. The rapidly moving ``point-like'' particles constantly collide with each other or with the walls of the container and exchange kinetic energy.  Below we describe in details, how simple simulations can be used to demonstrate the results of the kinetic theory of `ideal gases', which can be adapted to modelling simple closed economic systems for studying income/wealth distribution.

\subsection{Kinetic energy exchange model (`Ideal gas')}	
\begin{figure}
	\centering
		\includegraphics[width=0.7\linewidth]{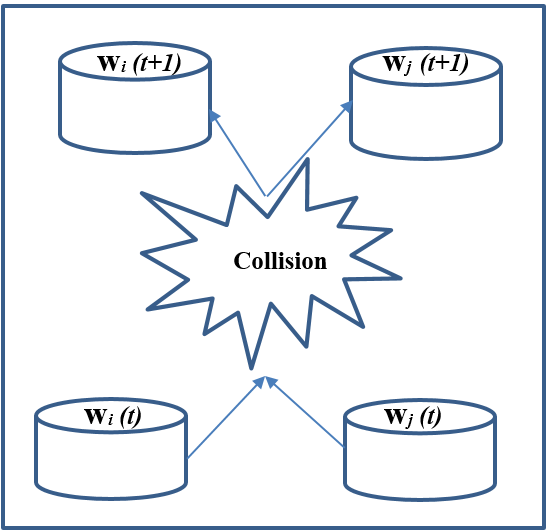}
\caption{
\textbf{Example of kinetic energy exchange.} Two particles $i$ and $j$ taking part in energy exchange (`collision') process. Particles $i$ and $j$ have energies $w_i(t)$ and $w_j(t)$ at time $t$. After collision, their energies become $w_i(t+1)$ and $w_j(t+1)$, respectively.
}
\label{Fig:collision}
\end{figure}

Kinetic exchange models are stochastic models, which are interpreted in terms of energy exchanges in gas molecules. The kinetic exchange model describes the dynamics at a microscopic level, based on pair-wise molecular collisions. Boltzmann wrote that `molecules are like so many individuals, having the most various states of motion' \cite{Boltzmann1872}. Thus, for two particles $i$ and $j$ with energies $w_i(t)$ and $w_j(t)$ at time $t$, the general dynamics can be described by the mathematical equations:
\begin{eqnarray}
w_i(t+1) = w_i(t) + \Delta w ; \nonumber \\
w_j(t+1) = w_j (t) - \Delta w , 
\label{Eq:energy_exchange}
\end{eqnarray}
\noindent where time $t$ changes by one unit after each collision. A typical energy exchange process is shown schematically in Fig. \ref{Fig:collision}.

\vspace{0.3cm}
\noindent\fbox{

\parbox[c]{\textwidth}{
The \textbf{Boltzmann-Gibbs} distribution \cite{wikiBoltzmann}, a fundamental law of equilibrium statistical mechanics, states that the probability $P(w)$ of finding a physical system or subsystem in a state with energy $w$ is given by the exponential function:
\begin{eqnarray}
 P(w) = Ce^{-w/T},  
 \label{eq:gibbs}
\end{eqnarray}
\noindent where $T$  is the temperature (average kinetic energy) of the system, $C$ is a constant and the conserved quantity is the total energy of the system.}
}
\vspace{0.3cm} 
 
\subsection{Simulation of the kinetic energy exchange model} 
Assume that the $N$ interacting units ${i}$, with $i =1,2,\dots,N$, are molecules of a gas with no interaction (potential) energy, and the variables ${w_i}$ represent their kinetic energies, such that $w_i \geq  0$. The time evolution of the system proceeds by a discrete stochastic dynamics \cite{Patriarca2013}. A series of updates of the kinetic energies $w_i(t)$ are made at the discrete times $ t = 0, 1 \dots $. Each update takes into account the effect of a collision between two molecules (as shown in the schematic diagram, Fig. \ref{Fig:collision}). The time step, which can be set to $ \Delta t = 1$ without loss of generality, represents the average time interval between two consecutive molecular collisions; i.e., on average, after each time step $\Delta t$, two molecules $i$ and $j$ undergo a `scattering' process and an update of their kinetic energies $w_i$ and $w_j$ is made. The evolution of the system is accomplished by the following steps at each time $t$:

\begin{enumerate}
\item	Randomly choose a pair of molecules $i$ and $j (i \neq  j)$ and $1 \leq  i,j \leq N$, with kinetic energies $w_i$ and $w_j,$, respectively; they represent the molecules undergoing a collision.

\item	Perform the energy exchange between molecules $i$ and $j$ by updating their kinetic energies,
\begin{eqnarray}
w_i(t+1) &=& r_t [w_i(t)+ w_j(t)], \nonumber \\
w_j(t+1) &=& (1- r_t) [w_i(t)+ w_j(t)],
\label{Eq:simulation}
\end{eqnarray}
\noindent where $r_t$ is a stochastic variable drawn as a uniform random number between $0$ and $1$, at time $t$. The total kinetic energy is conserved during an interaction.

\item	Increment the time step, and go to first step (see the \texttt{MATLAB} code given in the appendix).
\end{enumerate}

\begin{figure}
	\centering
		\includegraphics[width=0.7\linewidth]{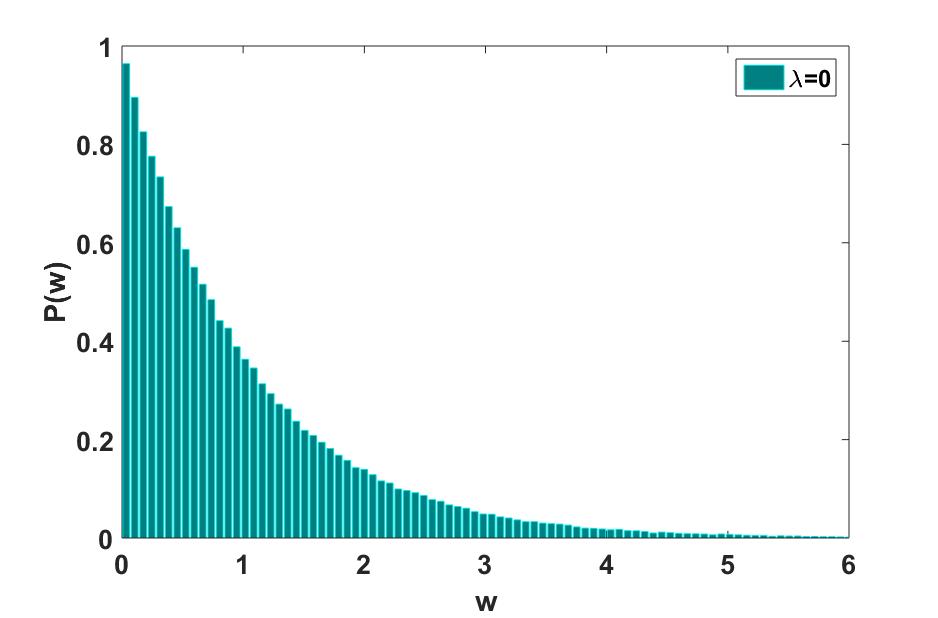}
\caption{
\textbf{Equilibrium energy distribution of the kinetic energy exchange model.} The results are obtained using the \texttt{MATLAB} code in the appendix, with parameters: $N=200$ particles and $T=50000$ time steps, and averaged over an ensemble of $k=2000$ runs. The distribution corresponds to the exponential Boltzmann-Gibbs distribution (Eq.~\ref{eq:gibbs}). The convergence to equilibrium is fast.}
\label{Fig:lambda0}
\end{figure}

For a large number of molecules ($N \rightarrow \infty $) and a sufficient number of time steps ($t \rightarrow \infty $), the system reaches an equilibrium (or steady-state) distribution \cite{Goswami2015}. The equilibrium distribution turns out to be the Boltzmann-Gibbs (exponential) distribution, as shown in Fig. \ref{Fig:lambda0}, which can be derived analytically in several ways-- probabilistic calculations \cite{Drăgulescu2000}, Master equation \cite{Chatterjee2005}, variational principle of maximum entropy \cite{Chakraborti2009}, etc. Interestingly, the most probable value of the exponential equilibrium distribution is zero (or very little) energy.

\section{Kinetic wealth exchange models}	
As mentioned in the introduction, understanding the distributions of income and wealth in an economy has been a classic problem in economics for more than a hundred years \cite{BKChakrabarti2013}. Inspired from the kinetic theory of gases (mentioned in the last section), the kinetic wealth exchange models (KWEMs) were proposed, which tried to explain the robust and universal features of income/wealth distributions. These form a class of simple multi-agent models, where the actions and interactions of autonomous agents (representing individuals, organizations, societies, etc.), could be used to understand the behaviour of the system as a whole \cite{Yakovenko2009,Chatterjee2007,Chatterjee2010,Chakraborti2002,Hayes2002,Lallouache2010,Matthes2008-09}.

KWEMs owe their popularity to the fact that they can capture many of the robust features of realistic wealth distributions using a minimal set of exchange rules \cite{Patriarca2013,Chakraborti2002,Patriarca2004}. In KWEMs, the closed economy or society is described in terms of a simple model, in which agents randomly meet and exchange a part of their wealth \cite{Patriarca2013}, similarly to particle assemblies (e.g., a gas) in which from time to time, a pair of particles collide and exchange energy, as given by Eqs. \ref{Eq:energy_exchange}. The core of the model dynamics are the simple linear relations in Eqs. \ref{Eq:energy_exchange}, and the difficulty is actually in generalizing and adapting the models, and solving analytically the equations. In fact, the exchange of wealth between two agents parallels the exchange of energy between colliding particles to the point that the kinetic theory of a gas in $D$ dimensions can suggest the expressions for the equilibrium distributions. 

\subsection{Model with no saving}	
The first model of this type was introduced by J. Angle in the context of social science (see Refs. \cite{Angle1986,Angle2006}), already some years earlier than in physics or economics. In the 1960's, Mandelbrot had suggested the possibility `. . . to consider the exchanges of money which occur in economic interaction as analogous to the exchanges of energy which occur in physical shocks between gas molecules ...' \cite{Mandelbrot1960}, but it was not until the works of E. Bennati \cite{Bennati1993} that such an analogy between statistical mechanics and the economics of wealth exchange was realized in terms of a quantitative Monte Carlo model, for which the corresponding numerical simulations demonstrated that the Boltzmann-Gibbs distribution was the equilibrium wealth distribution. Later, many physicists independently discovered such results by Monte Carlo simulations \cite{Drăgulescu2000,Ispolatov1998}. However, the introduction of `saving propensity' (\cite{Chakraborti2000}), as will be described next, brought forth the Gamma-like feature \cite{Patriarca2004,Chakraborti2008} of the distribution $P(w)$ and such a kinetic exchange model with uniform saving propensity for all agents was subsequently shown to be equivalent to a commodity clearing market, where each agent maximizes his/her own utility \cite{Chakrabarti2010-4}.

\subsection{Model with uniform saving}	
\begin{figure}
	\centering
		\includegraphics[width=0.48\linewidth]{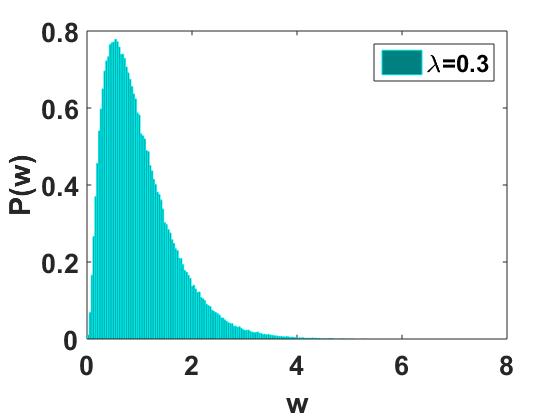}
		\llap{\parbox[b]{2.3in}{(a)\\\rule{0ex}{1.5in}}}
		\includegraphics[width=0.48\linewidth]{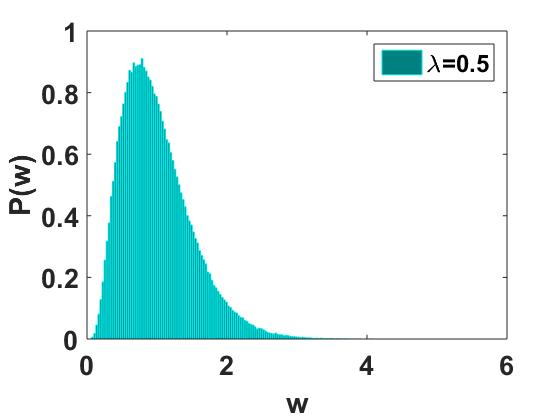}
		\llap{\parbox[b]{2.3in}{(b)\\\rule{0ex}{1.5in}}}
		\includegraphics[width=0.48\linewidth]{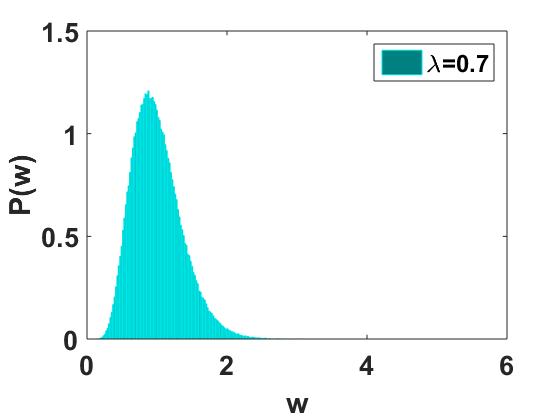}
		\llap{\parbox[b]{2.3in}{(c)\\\rule{0ex}{1.5in}}}
		\includegraphics[width=0.48\linewidth]{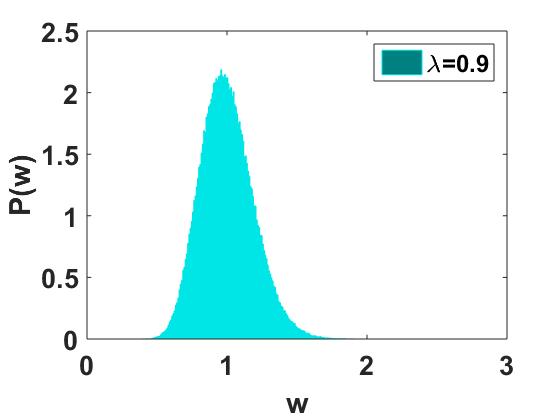}
		\llap{\parbox[b]{2.3in}{(d)\\\rule{0ex}{1.5in}}}
\caption{\textbf{Steady state wealth distribution with saving propensity $\lambda$.} Numerical simulations of the model defined by Eq.~(\ref{eq:lambda}) are best fitted by the Gamma-distribution (see \cite{Patriarca2004,Lallouache2010}). The results are obtained using the \texttt{MATLAB} code in the appendix, with parameters: $N=200$ particles and $T=50000$ time steps, and averaged over an ensemble of $k=2000$ runs. Panels (a)-(d) are for $\lambda=0.3, 0.5, 0.7, 0.9$, respectively. A direct inference that may be drawn from inspection of the figures is that as $\lambda$ increases, the inequality decreases, which has been studied extensively in Ref.~\cite{Sharma2018}.}
	\label{Fig:lambda}
\end{figure}

The concept of saving propensity was considered in this framework, first by Chakraborti and Chakrabarti \cite{Chakraborti2000}. In this model, the agents save a fixed fraction $ 0 \leq \lambda \leq 1$ of their wealth, when interacting with another agent. Thus, two agents with initial wealth $ w_i(t)$ and $w_j(t)$ at time $t$ interact such that they end up with wealth $ w_i(t+1)$ and  $w_j(t+1)$ given by
\begin{eqnarray}
w_i(t+1) &= & \lambda w_i(t) + r_t [(1 -  \lambda)( w_i(t) + w_j(t))], \nonumber \\
w_j(t+1) &= & \lambda w_j(t) +(1 – r_t)[(1 -  \lambda)( w_i(t) + w_j(t))],
\label{eq:lambda}
\end{eqnarray}
\noindent where $r_t$ is a stochastic fraction between $0$ and $1$, drawn from an uniform random distribution at time $t$.
If   $\lambda = 0$, equivalent to Bennati model \cite{Bennati1993}, then the most probable wealth per agent is zero, and the market is `non-interacting'. The market dynamics freezes (no interactions occur) when $\lambda = 1$. For the uniform saving propensity $\lambda$ in between the two limits, ($ 0 < \lambda < 1$), the steady state distribution $P(w)$ of money is a Gamma-like distribution \cite{Patriarca2004,Chakraborti2008,Lallouache2010} with exponentially decaying on both sides and the most-probable money per agent shifting away from $0$ (for $\lambda = 0$) to the average wealth of the system, $W/N$, as $\lambda \rightarrow 1$ (see Fig. \ref{Fig:lambda}). Note that there is no closed-form analytical solution of this exchange dynamics for the model with $ 0 < \lambda < 1$ and the Gamma distribution is the one which fits closest/best the simulation results (the first three moments matching exactly and the fourth moment differing \cite{Lallouache2010}). Here, the ``self-organizing'' feature of the market, simply induced by the ``self-interest'' of saving by each agent without any global perspective, is quite significant as the fraction of poor people (with very little or no money) decrease with saving propensity $\lambda$, and most people end up with some finite fraction of the average money in the market. The fact that savings can reduce inequality has been studied from a data science perspective in a recent article by Sharma et al. \cite{Sharma2018}. They also studied the empirical data of Gini indices and gross domestic savings (GDS) for several countries, and looked at the co-evolution of the countries in the inequality or savings spaces. Further, they  sought an empirical linkage
between the income inequality and savings, mainly for relatively
small or closed economies, using linear regression model.

Note also that $\lambda \rightarrow 1$ corresponds to the case where the economy is ideally `socialistic' (inequality of wealth is almost zero), and this is achieved just with the people's self-interest of saving. Although this fixed saving propensity does not give yet the Pareto-like power-law distribution, the Markovian nature of the scattering or trading processes is effectively lost. Indirectly through $\lambda$, the agents get to know (start `interacting' with) each other and the system co-operatively self-organizes towards a non-zero most-probable distribution (see Fig. \ref{Fig:lambda}). Thus, for $ 0 < \lambda < 1$, the market is effectively `interacting' \cite{Goswami2015}. The relaxation time to reach the steady state distribution is a complicated function of the saving propensity $\lambda$ and the system size $N$ \cite{Patriarca2007}.

Most interestingly for the physicists, the KWEMs in the regime $0 < \lambda < 1$, seem to reproduce the energy dynamics of a $D$-dimensional system, with the additional remarkable feature that the corresponding dimension $D$ can assume any real value \cite{Chakraborti2009,Goswami2015,Patriarca2004}, and also establishes interesting links with a generalized $D$-dimensional kinetic theory with real spatial dimension $ D$ \cite{Patriarca2015,Patriarca2017}. Numerical simulations suggest that at equilibrium, the system has a Gamma-distribution of order $n$, coinciding with the Boltzmann--Gibbs energy distribution for a $D$-dimensional gas with $D = 2n$. The relation between the dimension $D$ (or the order $n = D/2$) is: $D = 2(1 + 2 \lambda)/(1- \lambda)$. For $\lambda = 0$, the purely exponential shape is regained.

\subsection{Model with distributed savings}	
In a real society or economy, the interest of saving varies from person to person, which implies that $\lambda$ may be a heterogeneous parameter. To mimic this situation, the saving factor $\lambda$ may be assumed to be widely distributed within the population \cite{Chatterjee2004,Chatterjee2003}.
 
As before, starting with an arbitrary initial (uniform or random) distribution of wealth among the agents, the market evolves with the trading. At each time, two agents are randomly selected and the wealth exchange among them occurs, following the above mentioned scheme. One checks for the steady state, by looking at the stability of the money distribution in successive Monte Carlo steps $t$ (one Monte Carlo time step is defined as $N$ pairwise exchanges). Eventually, after a typical relaxation time the wealth distribution becomes stationary. This relaxation time is dependent on system size $N$ and the distribution of $\lambda$ (e.g., $~10^6$ for $N = 1000$ and uniformly distributed $\lambda$). After this, one averages the money distribution over $~10^3$ time steps. Finally, one takes the configurational average over $~10^5$ realizations of the $\lambda$ distribution to get the money distribution $P(w)$. Interestingly,  this non-ergodic model has a power-law decay similar to the decay of the Pareto law (see Eq. \ref{Eq:energy_dist}) with $\alpha \approx 1$. One may note, for finite size $N$ of the market, the distribution has a narrow initial growth up to a most-probable value, after which it decays as a power-law tail for several decades, and then there is a finite cut-off. This Pareto law (with $\alpha \approx 1$) covers almost the entire range in wealth $w$ of the distribution $P(w)$ in the limit $N \rightarrow \infty$. This result can be derived analytically too \cite{Mohanty2006,Chakraborti2009}.

\section{Discussions}	
There have been several works and extensions done on these simple KWEMs and lot of interesting features were extracted \cite{Chakraborti2002}. Recently, a `bi-directional exchange model` was introduced for mimicking more realistically a wealth exchange \cite{Heinsalu2014,Heinsalu2015}:
\begin{eqnarray}
w_i(t+1) &=& r_t w_i(t) + q_t w_j(t) ,		\nonumber	\\	
w_j(t+1) &=& (1-r_t)w_i(t) +(1-q_t)w_j(t), 
\end{eqnarray}
\noindent where $r_t$ and $q_t$ are two independent random numbers ($0,1$) and the sum of the variables before and after the collision is conserved, $w_i(t+1) + w_j(t+1) = w_i(t) + w_j(t)$. In the same paper \cite{Heinsalu2014}, a generalized microscopic version of the model was also introduced, in which, instead of saving propensity parameters, a few parameters regulated probabilistically the microscopic negotiation dynamics between the two agents. For this model, the numerical fitting of the equilibrium distributions suggested an effective dimension $D$ which is just half of the dimension of the saving propensity model of Ref. \cite{Chakraborti2000}, i.e., $D = (1+2 \lambda)/(1- \lambda$). This result, as well as the analogous ones based on the numerical fittings of the results of other versions of KWEMs, had remained more a conjecture for some years \cite{Lallouache2010,Patriarca2004,Repetowicz2005}. However, this bi-directional exchange model being more tractable analytically, Katriel \cite{Katriel2015} recently managed to show with the help of a Boltzmann equation approach that the relation between the dimension $D$ and the saving parameter $\lambda$ is exact, thus confirming the deep link between KWEMs and statistical physics. The study of this physics-related aspect of the models has now re-entered a challenging active phase in which the theoretical picture of the relaxation and equilibrium in KWEMs is under investigation.

The kinetic exchange framework can suitably be adapted to other areas in economics as well, e.g., the study of the firm size distributions. The size of a firm is measured by the strength of its workers. A firm grows when one worker joins it after leaving another firm. The rate at which a firm gains or loses workers is called the `turnover rate' in economics literature. Thus, there is a redistribution of workers and the corresponding dynamics can be studied using these exchange models. In the models of firm dynamics, one assumes that: (i) Any formal unemployment is avoided in the model. Thus one does not have to keep track of the mass of workers who are moving in and out of the employed workers pool. (ii) The number of workers is treated as a continuous variable. (iii)	The size of a firm is just the number of workers working in the firm.
In firm dynamics models, one makes an analogy with the previous subsections that firms are agents and the number of workers in the firm is its wealth. Assuming no migration, birth and death of workers, the economy thus remains conserved. As the `turnover rate' dictates both the inflow and outflow of workers, we need another parameter to describe only the outflow. That parameter may be termed as `retention rate', which describes the fraction of workers who decide to stay back in their firm. This is identical to saving propensity in wealth exchange models, as discussed earlier. Some interesting results using this framework have been produced by Chakrabarti \cite{Chakrabarti2012,Chakrabarti2013}.

Emergence of consensus is another important issue in sociophysics problems, where the people interact to select an option among different options of a subject like vote, language, culture, opinion, etc. \cite{Castellano2009,Galam2012,Stauffer2013,Sen2013}. When each person chooses an option, often a state of consensus is reached. In opinion formation, consensus is analogous to an `ordered phase' in statistical physics, where most of the people have a particular opinion. Several models have been proposed to mimic the dynamics of opinion spreading, and the opinions are usually modelled as discrete or continuous variables and are subject to either spontaneous changes or changes due to binary interactions, global feedback and external factors (see \cite{Castellano2009} for a general review). Lallouache et al. \cite{LallouacheI2010,LallouacheII2010} proposed a minimal multi-agent model for the collective dynamics of opinion formation in the society, by modifying kinetic exchange dynamics studied in the context of wealth distribution in a society. The model presented an intriguing spontaneous symmetry-breaking transition to polarized opinion state starting from non-polarized opinion state \cite{LallouacheII2010}, and many other features of interest to the statistical physicists studying phase transitions.

The most interesting aspect of KWEMs is related to their heterogeneous generalizations. In fact, KWEMs owe their popularity to the fact that they can predict realistic wealth distributions using a minimal set of exchange rules \cite{Patriarca2013,Chakraborti2002} but this cannot be achieved in the framework of homogeneous versions of the models, that (as mentioned earlier) usually lead to the exponential or Gamma-like equilibrium wealth distributions. Instead, the simple addition of a suitable level of heterogeneity, either in the saving propensities of the agents of the KWEM of Ref. \cite{Chakraborti2000}, or in the negotiation parameters of the model in Ref. \cite{Heinsalu2014} directly leads to the Pareto power-law. In other words, KWEMs suggest that heterogeneity is a key factor in producing the power-law observed in the wealth distributions \cite{Patriarca2005,Patriarca2010,Patriarca2015,Patriarca2017b}, a fact related to the general interest toward the effects of diversity in `Complex Systems' \cite{Patriarca2015}. Also, the dynamics of kinetic exchange models are often criticized for being based on an approach that is far from an actual economic or sociological foundation. However, it has been recently shown, for example, that such a economical dynamics of wealth exchange can also be derived from microeconomic theory \cite{Chakrabarti2010-4,Chakrabarti2009}. Although standard economics theory assumes that the activities of individual agents are driven solely by the utility maximization principle, the alternative picture that was presented, is that the agents can also be viewed as particles exchanging `wealth', instead of energy, and trading in wealth (energy) conserving two-body scattering, as in entropy maximization based on the kinetic theory of gases. This qualitative analogy between the two maximization principles has thus been firmly established only recently.

\section{Final remarks}	
Human beings are much more complex than particles!! The diverse types of interactions among the heterogeneous human beings make the society even more complex. However, in a certain idealised and simple closed economy (as mentioned in this article), we may ignore many complexities or ``degrees of freedom'' of the system, and model the system simply as an assembly of atoms or gas particles, in order to reproduce some of the statistical features of the empirical distributions. Definitely the real economy or society is much more complex, but this is perhaps one baby step of the econophysicists towards modelling the reality!

In this article, we could give an exposure to only a few models in one particular area. However, the field of Econophysics has had many many contributions from physicists, economists, mathematicians, financial engineers and others in the last two decades. Important directions and new areas in Econophysics have emerged in the last two decades, and one could get further information from the following Refs.:
\begin{itemize}
\item Empirical characterization, analyses and modelling of financial markets and limit order books \cite{ChakrabortiI2011,ChakrabortiII2011,Abergel2011,ChakrabortiI2016}.
\item Network models and characterization of market correlations among different stocks/sectors \cite{Abergel2012,Sharma2017a,Sharma2018b,Sharma2018c}.
\item Determination of the income or wealth distribution in societies, and the development of statistical physics models \cite{Chakrabarti2005,BKChakrabarti2013}.
\item Development of behavioural models, and analyses of market bubbles and crashes \cite{Abergel2013,Abergel2015,Pharasi2018}.
\item Learning in multi-agent game models and the development of Minority Game models \cite{Chakraborti2015,ChakrabortiI2015,Sharma2018d}.
\end{itemize}  

\section{Acknowledgment}	
This manuscript was based on the lectures given by AC at International Christian University-Mitaka, Japan, the Indian Institute of Advanced Study (IIAS)-Shimla, India and the International Conference on ‘Social Statistics in India’ at Asian Development Research Institute-Patna, India.
AC and KS heartily thank all the collaborators whose works have been represented here. They also acknowledge the hospitality of IIAS-Shimla where the manuscript was initiated. KS thanks University Grants Commission (Ministry of Human Research Development, Govt. of India)  for her senior research fellowship. AC acknowledges financial support from the grant number BT/BI/03/004/2003(C) of Govt. of India, Ministry of Science and Technology, Department of Biotechnology, Bioinformatics division; University of Potential Excellence-II grant (Project ID-47) of the Jawaharlal Nehru University, New Delhi; the DST-PURSE grant given to Jawaharlal Nehru University, New Delhi.

\section{Appendix}
Code in \texttt{MATLAB} for generating the equilibrium distributions represented in Fig. \ref{Fig:lambda0}:

 \begin{framed}
 \includegraphics[width=0.95\linewidth]{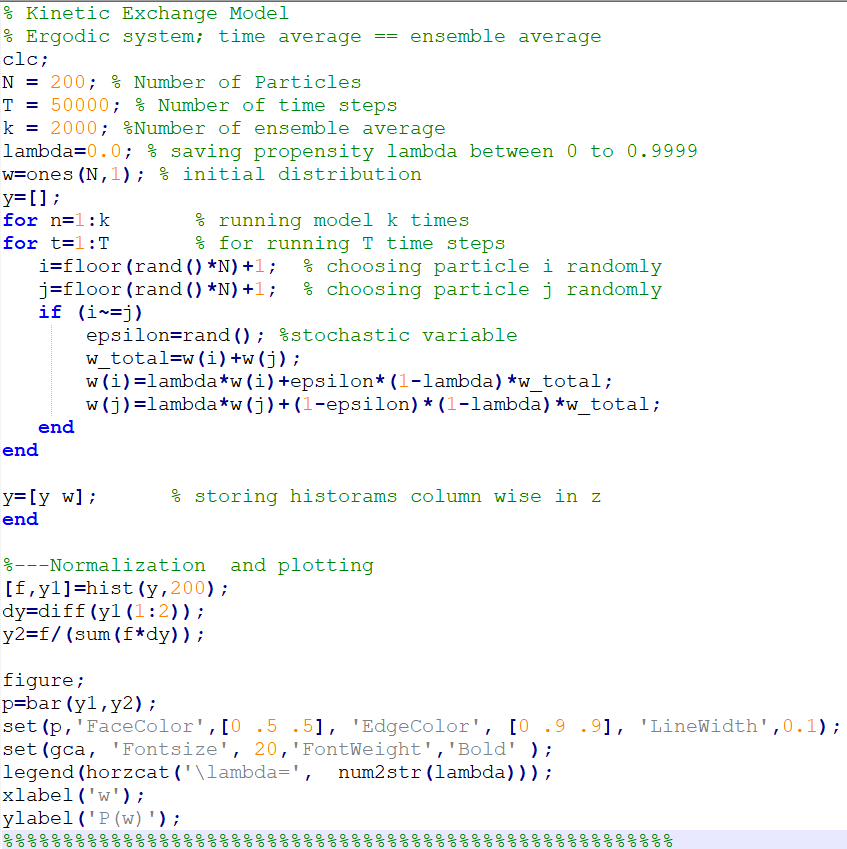}%
 \end{framed}

%
%

\end{document}